\documentclass[10pt,aps,prd,twocolumn,
nofootinbib,preprintnumbers,superscriptaddress,showpacs]{revtex4-1}

\usepackage{times}
\usepackage{amssymb,amsbsy,amsmath,amsfonts}
\usepackage{mathrsfs}
\usepackage{graphicx}% Include figure files
\usepackage{graphics,psfrag,longtable}
\usepackage{nicefrac}
\usepackage{esint}
\usepackage{upgreek}
\usepackage{multirow}
\usepackage{rotating}
\usepackage{diagbox}

\usepackage[colorlinks,linktoc=all,citecolor=blue,linkcolor=red]{hyperref}

\def\beq{\begin{equation}}
\def\eeq{\end{equation}}
\def\bea{\begin{eqnarray}}
\def\eea{\end{eqnarray}}
\def\beqa{\begin{equation}\begin{array}{l}}
\def\eeqa{\end{array}\end{equation}}

\def\eqlab#1{\label{eq:#1}}
\def\figlab#1{\label{fig:#1}}

\def\seclab#1{\label{sec:#1}}
\def\eref#1{(\ref{eq:#1})}
\def\eqref#1{eq.~(\ref{eq:#1})}
\def\Eqref#1{Eq.~(\ref{eq:#1})}

\def\Figref#1{Fig.~\ref{fig:#1}}

\def\half{\mbox{\small{$\frac{1}{2}$}}}

\def\barr{\left(\begin{array}{c}}
\def\earr{\end{array}\right)}
\def\bmat{\left(\begin{array}{cc}}
\def\emat{\end{array}\right)}
\def\al{\alpha}

\def\ga{\gamma} 
 \def\De{\Delta}
  \def\eps{\epsilon}

\def\la{\lambda}

\def\si{\sigma}

\def\dd{{\rm d}}

\def\nn{\nonumber}

\DeclareMathOperator\im{Im}
\DeclareMathOperator\re{Re}

\def\beps{\boldsymbol{\varepsilon}}

\def\bsigma{\boldsymbol{\sigma}}

\begin{document}
\preprint{MITP/15-018}

\title {Evaluation of the forward Compton scattering off protons: I. Spin-independent amplitude}
\author{Oleksii Gryniuk}
\affiliation{Institut f\"ur Kernphysik \& PRISMA Cluster of Excellence, Johannes
Gutenberg-Universit\"at Mainz, D-55128 Mainz, Germany}
\affiliation{Physics Department, Taras Shevchenko Kyiv National University, Volodymyrska 60, UA-01033 Kyiv, Ukraine}
\author{Franziska Hagelstein}
\author{Vladimir Pascalutsa}
\affiliation{Institut f\"ur Kernphysik \& PRISMA Cluster of Excellence, Johannes
Gutenberg-Universit\"at Mainz, D-55128 Mainz, Germany}

\begin{abstract}
We evaluate the forward Compton scattering off the proton, based on Kramers-Kronig kind of relations
which express the Compton amplitudes in terms of integrals of total photoabsorption cross sections.
We obtain two distinct fits to the world data on the unpolarized total photoabsorption 
cross section, and evaluate the various spin-independent sum rules using these fits. For the sum 
of proton electric and magnetic dipole polarizabilities, governed by the Baldin sum rule, we obtain the following average (between the two fits): $\al_{E1}+\beta_{M1}=14.0(2)\times 10^{-4}\,\mathrm{fm}^3$. An analogous sum rule involving the quadrupole polarizabilities of the proton is
evaluated too. The spin-independent forward amplitude of proton Compton scattering is
evaluated in a broad energy range. The results are compared with previous evaluations and 
the only experimental data point for this amplitude (at 2.2 GeV). 
We remark on sum rules for the elastic component of polarizabilities.
\end{abstract}
\pacs{13.60.Fz - Elastic and Compton scattering,
14.20.Dh - Protons and neutrons,
25.20.Dc - Photon absorption and scattering,
11.55.Hx Sum rules}
\date{\today}
\maketitle

%\tableofcontents

\section{Introduction}
It  is long known that  the forward Compton scattering (CS) amplitudes can, by unitarity, causality and crossing, be expressed
through integrals of the photoabsorption cross sections~\cite{GellMann:1954db}. 
The low-energy expansions of these expressions lead to a number of useful sum rules, most notably 
those of Baldin \cite{Baldin}, and Gerasimov, Drell and Hearn (GDH)~\cite{Gerasimov:1965et,Drell:1966jv}. Given the photoabsorption cross sections,
one can thus provide a reliable assessment of some of the static electromagnetic 
properties of the nucleon and nuclei, as well as of the forward CS amplitudes in general.
For the proton, the first such assessments was performed in the early 
1970s \cite{Damashek,Armstrong}. Since then, the knowledge of the photoabsorption cross sections
appreciably improved, and yet for the unpolarized case only the Baldin
sum rule has been updated~\cite{Schroder,Babusci,Olmos}. 
In this work, we provide a re-assessment of the forward spin-independent amplitude of proton CS, and evaluate the associated sum rules involving the
{\em dipole and quadrupole polarizabilities of the proton}.

Sum rules are essentially the only way to gain empirical knowledge of 
the forward CS amplitudes.  It is impossible to access the forward kinematics 
in real CS experiments. The measurement of the forward spin-independent CS amplitude can be done indirectly through the process of dilepton photoproduction ($\gamma \,p \to  p \,e^+ e^-$)~\cite{Alvensleben}. The timelike CS, involved
in the process of dilepton photoproduction, yields access to real CS given 
the small virtuality of the outgoing photon, or equivalently,  the nearly vanishing invariant
mass of the produced pair. The experimental result~\cite{Alvensleben} 
compared well with the aforementioned evaluations \cite{Damashek,Armstrong}. 
Despite the substantial
additions to the database of total photoabsorption cross sections, the works of Damashek and Gilman (DG) \cite{Damashek} as well as Armstrong {\it et al.}~\cite{Armstrong} remained to be, until now, the only evaluations
of the full amplitude.

The newer data were used, however, in the most recent evaluations of the Baldin sum rule \cite{Babusci,Olmos}, which 
yields the sum of the electric and magnetic dipole polarizabilities,
\Eqref{BaldinSR}. These recent analyses obtained somewhat lower value for the sum than DG, cf.~Table \ref{sumruletest}. In this work we find that the difference between the early and the recent evaluations arises from systematic inconsistencies in the experimental database. 
We also obtain the sum rule value for a combination of higher-order quadrupole polarizabilities and 
compare it with several theoretical predictions.

This paper is organized as follows.
In Sect.\ \ref{sec:overview} we give a brief overview of 
the Kramers-Kronig relation and sum rules for polarizabilities.
In Sect.\ \ref{sec:fitting} we discuss the fitting procedure
for the unpolarized total proton photoabsorption cross section data.
The sum rule evaluations of scalar polarizabilities
and of the spin-independent forward CS amplitude are presented 
in Sect.\ \ref{sec:sumrules}.
Conclusions are given in Sect.\ \ref{sec:summary}. The Appendix
demonstrates the elastic-channel contribution to the sum rules and polarizabilities on the example of one-loop scalar QED.

\begin{figure*}[hbt]
\includegraphics[width=0.8\textwidth]{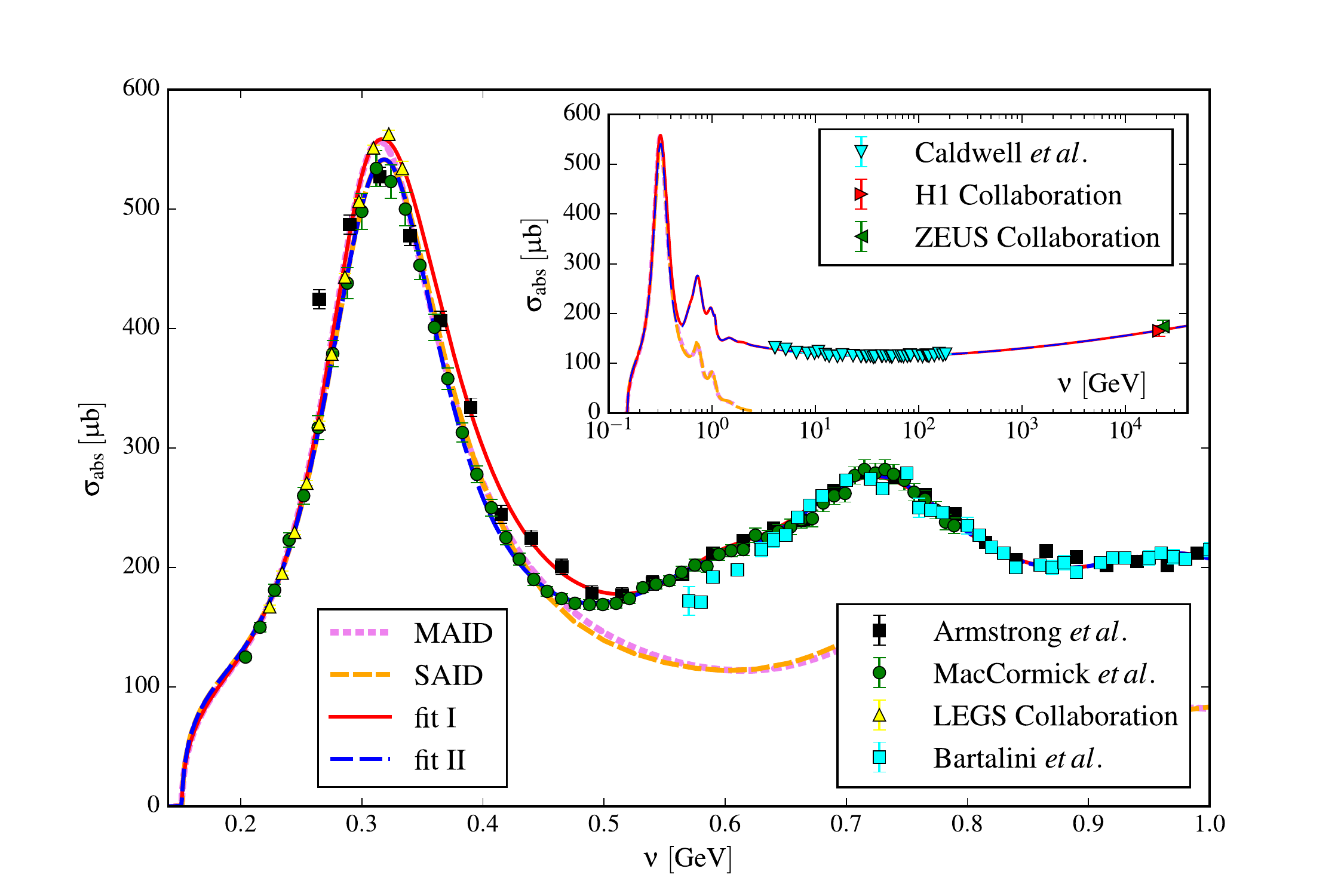}
\caption{
(Color online)
Fits of experimental data for the total photoabsorption cross section on the proton.  Fit I is obtained using MAID \cite{MAID} results below the $2\pi$-production, and data from LEGS \cite{BNL} and Armstrong {\it et al.}~\cite{Armstrong} above it. Fit II uses SAID \cite{SAID} and the data of MacCormick {\it et al.}~\cite{MacCormick}. Both fits use Bartalini {\it et al.} \cite{Bartalini} and the high-energy data \cite{Caldwell,Aid,Chekanov} displayed in the insert.
}
\label{fig:CS_tot}
\end{figure*}
\section{Forward Compton Amplitude and Sum Rules}
\label{sec:overview}

For a spin-1/2 target, such as the proton, 
the forward CS amplitude is given by
\beq 
T_{fi} = f(\nu) \, \beps^{\prime\,\ast }\cdot \beps \, + \, g(\nu) \,i \, (\beps^{\prime\,\ast }\! \times \beps)\cdot\bsigma,
\eqlab{definition}
\eeq 
where $f$ and $g$ are scalar functions of the photon lab energy $\nu$;
%,are the spin-independent and spin-dependent amplitudes
vectors $\beps$ and $\bsigma$ represent the photon and proton polarizations, respectively. The crossing symmetry implies that the spin-independent amplitude $f$ is an even
and the spin-dependent amplitude $g$ is an odd function of $\nu$.

The optical theorem (unitarity) relates the imaginary
part of the amplitudes to the total photoabsorption cross sections:
\begin{subequations}
\begin{eqnarray}
\im f(\nu) & = & \frac{\nu}{8\pi} \left[\sigma_{1/2}(\nu)+\sigma_{3/2}(\nu)\right] , \\
%%%%%%%%%%%%
\im g(\nu) & = & \frac{\nu}{8\pi}
\left[\sigma_{1/2}(\nu)-\sigma_{3/2}(\nu)\right] .
\eqlab{unitarity}
\end{eqnarray}
\end{subequations}
Here $\sigma_{\la}(\nu) $ is the doubly-polarized cross section
with $\lambda$ representing the
combined helicity of the initial $\ga p$ state.  Averaging over the
polarization of initial particles gives the  unpolarized
photoabsorption cross section, $\sigma =\half ( \sigma_{1/2}+\sigma_{3/2})$.

In the present article we focus on relations involving  the spin-independent amplitude $f$, and the unpolarized cross section
$\si$.
The Kramers-Kronig relation between these quantities exploits the optical theorem, causality and crossing symmetry, to yield for the proton~\cite{GellMann:1954db}:
\beq
\eqlab{KK}
\re f(\nu) =-\,\frac{\alpha}{M_p} + \frac{\nu^2}{2\pi^2}\fint_0^\infty\! \frac{\dd \nu'}{\nu^{\prime \, 2}-\nu^2}\, \si(\nu'),\eeq
where $\al=e^2/4\pi$ is the fine-structure constant and $M_p$ is the proton mass; the slashed integral denotes the principal-value integration. 

We next would like to consider the low-energy expansion of $f$.
At this point it is important to note that the elastic scattering, i.e.\ the CS process itself, 
is one of the photoabsorption processes. The total CS cross section does
not vanish for $\nu\to 0$ but goes to a constant --- the Thomson cross section: 
\beq 
\si(0 ) = \frac{8\pi \al^2}{3 M_p^2}.
\eeq 
This means \Eqref{KK} does not admit a Taylor-series expansion
around $\nu=0$ (each coefficient in that expansion is infrared divergent, cf. Appendix).
Such expansion is nonetheless important for establishing the polarizability sum rules.
We hence prefer to take the CS out of the total cross section, i.e.:
\beq 
\si(\nu) = \si_{\mathrm{CS}}(\nu)+\si_{\mathrm{abs}}(\nu), 
\eeq 
where $\si_{\mathrm{abs}}$ can be assumed to be dominated
by hadron-production processes, for which there is a threshold 
at some $\nu_0 > m_\pi$.

The amplitude $f$ can be decomposed accordingly into the elastic
and inelastic terms,
\bea 
f(\nu) & = & f_{\mathrm{el}}(\nu) + f_{\mathrm{inel}}(\nu)  , \\
f_{\mathrm{el}}(\nu) &=& -\,\frac{\alpha}{M_p} + \frac{\nu^2}{2\pi^2}\fint_0^\infty\! \frac{\dd \nu'}{\nu^{\prime \, 2}-\nu^2}\, \si_{\mathrm{CS}}(\nu'),
\eqlab{fel}\\
f_{\mathrm{inel}}(\nu) &=& \frac{\nu^2}{2\pi^2}\fint_{\nu_0}^\infty\! \frac{\dd \nu'}{\nu^{\prime \, 2}-\nu^2}\, \si_{\mathrm{abs}}(\nu').
\eqlab{fbar}
\eea 
The details on dealing with $f_{\mathrm{el}}$ can be found in 
Appendix \ref{sec:scalarQEDExample}. In what follows, however, we neglect
the contribution from $\si_{\mathrm{CS}}$, as it is suppressed by an extra order of $\al$. Hence we set $f_{\mathrm{el}}(\nu) = -\alpha/M_p$, as is usually done. 

Considering $f_{\mathrm{inel}}$,
the low-energy expansion of both sides of \Eqref{fbar} leads to the sum rules for polarizabilities. At the leading order [$O(\nu^2)$], one
obtains the Baldin sum rule \cite{Baldin} for the
sum of electric ($\alpha_{E1}$) and magnetic ($\beta_{M1}$) dipole polarizabilities:
\beq
\eqlab{BaldinSR}
\alpha_{E1} + \beta_{M1} =\frac{1}{2\pi^2} \int_{\nu_0}^\infty \!\dd\nu \,\frac{\si_{\mathrm{abs}}(\nu)}{\nu^2}.
\eeq
At $O(\nu^4)$ we obtain `the 4\textsuperscript{th}-order sum rule': 
\bea
\eqlab{4thSR}
\alpha_{E\nu} + \beta_{M\nu} + \frac{1}{12}\,(\alpha_{E2} + \beta_{M2}) =\frac{1}{2\pi^2} \int_{\nu_0}^\infty \!\dd\nu \,\frac{\si_{\mathrm{abs}}(\nu)}{\nu^4},\nn\\
\eea
which involves the quadrupole  polarizabilities
$\alpha_{E2}$, $\beta_{M2}$, as well as the leading dispersive
contribution to the dipole polarizabilities denoted as $\alpha_{E\nu}$, $\beta_{M\nu}$, see \cite{Babusci:1998ww} for more details.

Our aim here is to provide an empirical fit of the available
data for $\si_{\mathrm{abs}}$ and evaluate the various sum rules. 

\section{Fits of the Photoabsorption Cross Section}
\label{sec:fitting}

The presently available experimental data, together with
the results of the empirical analyses MAID and SAID, as well as our fits are displayed in \Figref{CS_tot}. In our fitting we
distinguish the following 
three regions:
\begin{itemize}
 \item {\it low-energy},   $\nu \in [\nu_0   \,,\;  \nu_1  ) $;
 \item {\it medium-energy},   $\nu \in [\nu_1   \,,\;  2\;\mathrm{GeV}) $;
 \item {\it high-energy}, $\nu \in [2\;\mathrm{GeV} \,,\;  \infty ) $;
\end{itemize}
where $\nu_0$ ($\simeq 0.145$ GeV) and $\nu_1$ ($\simeq 0.309$ GeV) are respectively the thresholds for the single- and double-pion photoproduction on the proton.

In the {\it low-energy} region we use the pion production 
($\pi^+  n$ and $\pi^0 p$) cross sections from the 
MAID \cite{MAID} and SAID \cite{SAID} partial-wave analyses. 
In our error estimate we assign a 2\% uncertainty on these values.

In the {\it medium-energy} region we fit the actual experimental data, using a sum of Breit-Wigner resonances and a background. 
Following \cite{Babusci}, we take 
six Breit-Wigner resonances, each parameterized 
as
\beq
\label{res_fit}
\si_R(W)=A \cdot \frac{\Gamma^2/4}{(W-M)^2 + \Gamma^2/4},
\eeq
where $W=\sqrt{s}$ is the total energy of the $\gamma p$ system.
The background function is from \cite{Armstrong}:
\beq
\label{bgd_fit}
\si_B(W)=\sum_{k=-2}^2 C_k (W-W_0)^k,
\eeq
where $W_0=M_p+m_{\pi}$ corresponds with
the pion photoproduction threshold.

\begin{table}[bth]
{
\begin{tabular}{|c||c|c|c|}
\hline
&$M$ [MeV] & $\Gamma$ [MeV] & $A$ [$\upmu$b]\\
\hline
\hline
\multirow{6}{*}{\begin{sideways} \hspace{-0.4cm} fit I\end{sideways}}&$1213.6 \pm  0.1$ & $117.6 \pm  1.9$ &  $522.7 \pm  17.0$\\
&$1412.8 \pm  5.9$ & $ 82.8 \pm 26.8$ &  $ 40.1 \pm 33.8$\\
&$1496.0 \pm  2.8$ & $136.5 \pm 11.1$ &  $161.8 \pm 32.4$\\
&$1649.4 \pm  4.1$ & $135.3 \pm 15.3$ &  $ 83.2 \pm 22.7$\\
&$1697.5 \pm  2.6$ & $ 18.8 \pm 12.6$ &  $ 18.2 \pm 26.0$\\
&$1894.3 \pm 15.6$ & $302.0 \pm 41.3$ &  $ 31.5 \pm  8.7$\\
\hline
\hline
\multirow{6}{*}{\begin{sideways}\hspace{-0.4cm}  fit II\end{sideways}}&$1214.8 \pm  0.1$ & $99.0 \pm  1.1$ & $502.3 \pm 12.3$\\
&$1403.9 \pm  6.2$ & $118.2 \pm 19.6$ & $ 51.8 \pm 23.8$\\
&$1496.9 \pm  2.1$ & $133.4 \pm  9.4$ & $162.0 \pm 29.2$\\
&$1648.0 \pm  4.4$ & $135.2 \pm 15.9$ & $ 83.6 \pm 23.8$\\
&$1697.2 \pm  2.7$ & $ 21.2 \pm 13.2$ & $ 18.7 \pm 25.9$\\
&$1893.7 \pm 17.4$ & $323.5 \pm 45.3$ & $ 31.7 \pm  9.1$\\
\hline
\end{tabular}
}
\caption{Fitting parameters for the resonances (\ref{res_fit})
obtained for fit I and II.}
\label{table:res_par_2}
\end{table}

\begin{table}[bth]
{
\begin{tabular}{|c|c|c|}
\hline
& fit I & fit II \\
\hline
$C_{-2}$, [$\upmu$b $\cdot$ GeV$^2$] & $ 0.44  \pm  0.22$ & $ 0.26 \pm  0.17$\\
$C_{-1}$, [$\upmu$b $\cdot$ GeV]     & $-11.06 \pm  3.69$ & $-7.97 \pm  2.89$\\
$C_0$, [$\upmu$b]                    & $ 74.38 \pm 20.16$ & $57.27 \pm 16.09$\\
$C_1$, [$\upmu$b $\cdot$ GeV$^{-1}$] & $ 22.18 \pm 37.71$ & $54.26 \pm 31.07$\\
$C_2$, [$\upmu$b $\cdot$ GeV$^{-2}$] & $ 37.69 \pm 21.48$ & $19.51 \pm 18.17$\\
\hline
\end{tabular}
}
\caption{Fitting parameters for the background (\ref{bgd_fit})
obtained for fit I and II in the resonance region.}
\label{table:bgd_par}
\end{table}

Observing a significant discrepancy between SAID and MAID 
around the $\De$(1232)-resonance peak and a similar discrepancy
between two sets of experimental data, we have made two
different fits:
\begin{itemize}
\item[I.] MAID\cite{MAID} + LEGS \cite{BNL} + Armstrong {\it et al.}~\cite{Armstrong}, 
\item[II.] SAID \cite{SAID} + MacCormick {\it et al.}~\cite{MacCormick}.
\end{itemize}
They are shown in \Figref{CS_tot} by the red solid and blue dashed lines, respectively. The corresponding values of parameters are
given in Tables~\ref{table:res_par_2} and \ref{table:bgd_par}.
In both fits we have also made
use of the GRAAL 2007 data \cite{Bartalini}, shown in the figure by
light-blue squares. These data had not been available at the time of the previous sum rule evaluations.

Finally, for the {\it high-energy region} we use the
standard Regge form \cite[p.~191]{Barnett}:
\beq
\label{regge_fit}
\si_\mathrm{Regge}(W)=c_1 \, W^{p_1} + c_2 \, W^{p_2}.\\
\eeq
For $W$ in GeV and the cross section in $\upmu$b, we obtain
the following parameters (for both of our fits):
\begin{gather*}
c_1 = 62.0 \pm 8.1, \quad c_2 = 126.3 \pm 4.3, \\
p_1 = 0.184 \pm 0.032, \quad p_2 = -0.81 \pm 0.12.
\end{gather*}
We also tried the high-energy parameterization used
in \cite{Babusci}, but obtained a worse fit and abandoned it.

\begin{figure*}[bht]
	\includegraphics[width=0.8\textwidth]{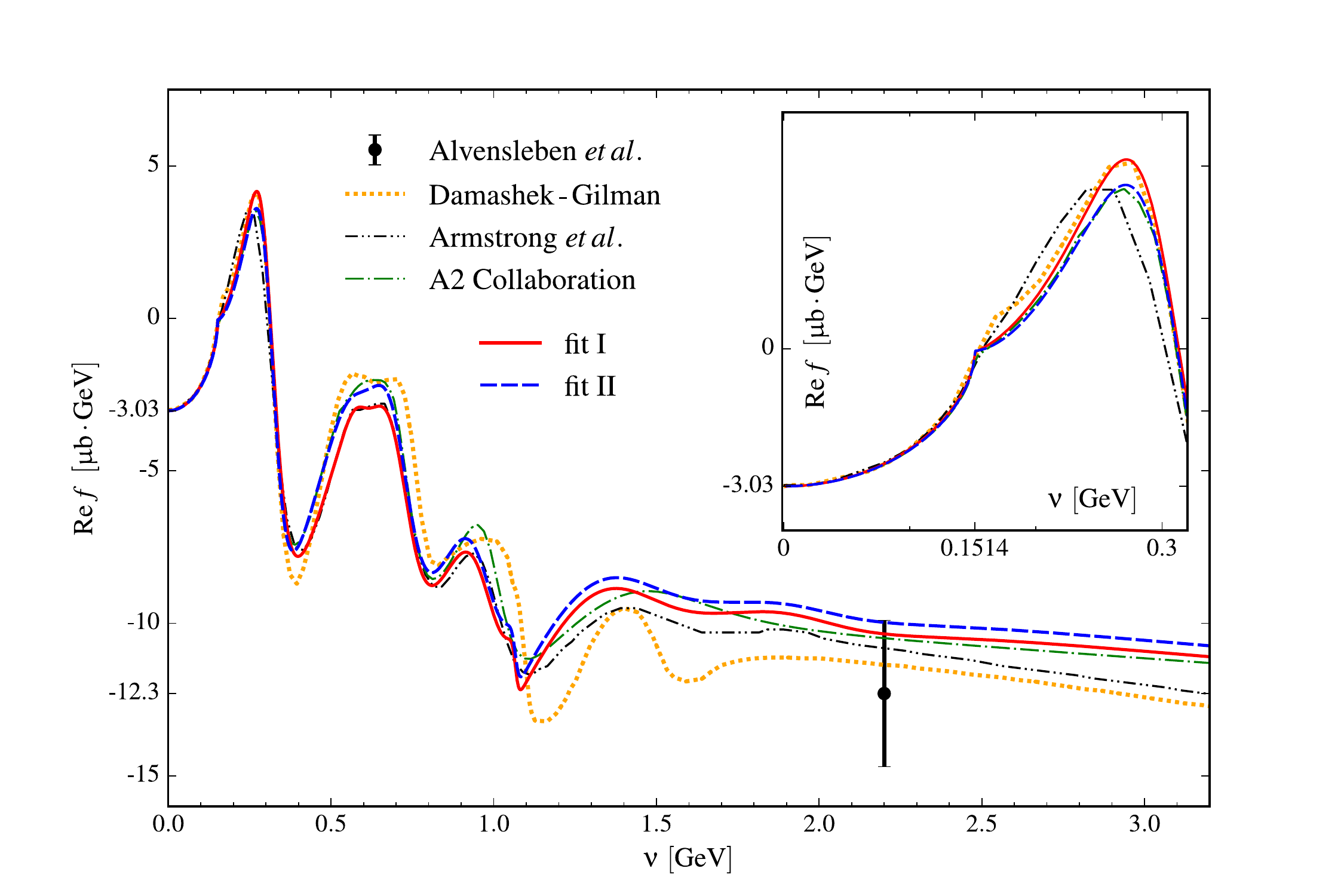}
	\caption{(Color online) Our evaluation of $\re f$ based on the two fits
of the photoabsorption cross section, compared with previous evaluations \cite{Damashek,Olmos,Armstrong}. The experimental data point is from Ref.~\cite{Alvensleben}.}
	\label{fig:Re_f}
\end{figure*}

\begin{table*}[tbh]
\begin{ruledtabular}
\centering 
\caption{
	Empirical evaluations of sum rules 
    and verification of the
    Kramers-Kronig relation for the proton.}
    \begin{tabular}{|r|l|l|l|l|}
	&  Baldin& 4\textsuperscript{th} order &  6\textsuperscript{th} order\footnote{$ \int_{\nu_0}^\infty \!\dd\nu\, \nu^{-6} \,\si_{\mathrm{abs}}(\nu)/(2\pi^2)$} & $\re f$($2.2$ GeV) \\
    	&[$10^{-4}$ fm$^3$]&[$10^{-4}$ fm$^5$]&[$10^{-4}$ fm$^7$]&[$\upmu$b $\cdot$ GeV]\\
	\hline
	Damashek--Gilman \cite{Damashek}      & $14.2\pm 0.3$  &  &  & $-11.5$\footnote{Interpolated value.}\\
    Armstrong {\it et al.} \cite{Armstrong} &              &  &  & $-10.8$\\
	Schr{\"o}der \cite{Schroder}            & $14.7\pm0.7$  &$6.4$&  & \\
	Babusci {\it et al.} \cite{Babusci}     & $13.69\pm0.14$ &  &  & \\
	A2 Collaboration \cite{Olmos} & $13.8\pm0.4$   &  &  & $-10.5$\footnote{Based on the cross-section parametrization from \cite{Ahrens}.}\\
	MAID ($\pi$ chan.)~\cite{MAID}  & $11.63$\footnote{Integrated from threshold to $\nu_\mathrm{max}=1.663$ GeV.}     &  &  & \\
	SAID ($\pi$ chan.)~\cite{SAID}             & $11.5$\footnote{Integrated from threshold to $\nu_\mathrm{max}=2$ GeV.}           &  &  & \\
	\hline
    This work &&&&\\
    	 fit I & $14.29\pm 0.27$ & $6.08 \pm 0.12 $ & $ 4.36 \pm 0.09 $ & $-10.35$ \\
	 fit II  & $13.85\pm 0.22$ & $6.01 \pm 0.11 $ & $ 4.42 \pm 0.08 $ & $-9.97 $ \\
     	\hline
    Experiment &&&&\\
      Alvensleben {\it et al.} \cite{Alvensleben} &          &  &  & $-12.3\pm2.4$\\
	\end{tabular}
	\label{sumruletest}
   \end{ruledtabular}
\end{table*}

\twocolumngrid 

The fitting was done with the help of the {\it SciPy} package for Python.
The resulting chi-square, evaluated as
\beq
\eqlab{chisq}
\chi^2 = \sum_i \frac{\left(\sigma_i^{\mathrm{fit}}  - \sigma_i^{\mathrm{exp}}\right)^2}{(\De\sigma_i^{\mathrm{exp}})^2},
\eeq
is of about the same quality for the two fits. In the intermediate region,
we obtain $\chi^2/\mbox{point}=0.7$ for fit I, and $\chi^2/\mbox{point}=0.6$
for fit II. In the high-energy region $\chi^2/\mbox{point}=1.2$ in both cases.
Again, the low-energy region is not fitted but is borrowed from, respectively, the MAID and SAID
analyses.

\section{Sum Rule Evaluations}
\label{sec:sumrules}

Having obtained the fits of the total photoabsorption cross section $\si_{\mathrm{abs}}$, we evaluate the integrals 
in Eqs.~\eref{fbar}, \eref{BaldinSR} and \eref{4thSR}; the
results are presented in \Figref{Re_f} and Table~\ref{sumruletest}.

Tables \ref{BaldinSR_regions} and \ref{4thSR_regions} show contributions of each region to the Baldin and the 4\textsuperscript{th}-order sum rule, respectively. 
The uncertainty in calculating an integral
$I_n = \int\dd\nu\,  \nu^{-n} \sigma(\nu) $ has been evaluated as follows:
\beq
\Delta I_n = \sum_i \frac{\Delta \nu_i}{\nu_i^n} \chi_i^2 \,\Delta\sigma_i^{\mathrm{exp}},
\eeq
where $\chi_i^2$ is the chi-square at the point $i$, cf.~\Eqref{chisq}.

\begin{table}[tbh]{
\caption{Contributions of different regions to the Baldin sum rule
for the two fits in \Figref{CS_tot}.}
   \label{BaldinSR_regions}
$\alpha_{E1} + \beta_{M1}$\; [$10^{-4}$ fm$^3$]\\
	\centering 
	\begin{tabular}{|c|c|c|c|}
	\hline
\diagbox{Fit}{Region}& {\it low-energy} & {\it medium-energy}  & {\it high-energy}  \\
	\hline
	I & $6.12 \pm 0.12$ & $7.53 \pm 0.13$ & $0.64 \pm 0.02$ \\
	II& $6.06 \pm 0.12$ & $7.15 \pm 0.08$ & $0.64 \pm 0.02$ \\
	\hline
	\end{tabular}
	\par}
 
\end{table}

\begin{table}[tbh]{
\caption{ Contributions of different regions to the 4\textsuperscript{th}-order sum rule
for the two fits in \Figref{CS_tot}.} \label{4thSR_regions}
$\alpha_{E\nu} + \beta_{M\nu} + \frac{1}{12}(\alpha_{E2} + \beta_{M2})$ \; [$10^{-4}$ fm$^5$]\\
	\centering 
	\begin{tabular}{|c|c|c|c|}
	\hline
	\diagbox{Fit}{Region} & {\it low-energy} & {\it medium-energy}   & {\it high-energy}  \\
	\hline
	I  & $4.50 \pm 0.09$ & $1.58 \pm 0.03$ & $(219 \pm 8)\times 10^{-5}$ \\
	II  & $4.53 \pm 0.09$ & $1.48 \pm 0.01$ & $(219 \pm 8)\times 10^{-5}$ \\
	\hline
	\end{tabular}
	\par}
\end{table}

The corresponding full results (sum of the three regions) 
are given in 
Table~\ref{sumruletest}, and compared with the results of previous works. In this table we also give the result for the 6th-order
integral, and for the full amplitude $f$ at $\nu=2.2$ GeV.
The real part of $f$ is plotted in \Figref{Re_f} over a broad
energy range and compared with previous evaluations and 
the experimental number from the 1973 DESY experiment at 
2.2 GeV. Although none of the evaluations really contradicts the experiment, 
there is a clear tendency to a higher central value.

The new
dilepton photoproduction experiments planned  at the Mainz Microtron (MAMI) could perhaps provide experimental values in the lower energy range. Obviously, the regions of the extrema (e.g., the $\De(1232)$ region or the interval between 0.6 and 0.7 GeV) are most interesting as the different evaluations seem to differ there the most. In
the region around 0.6 GeV, for example, one of our evaluations (fit I) is nearly 
identical with Armstrong's~\cite{Armstrong}, while the other one (fit II) is aligning with DG~\cite{Damashek} and A2 Coll.~\cite{Olmos}. An appropriately precise
experiment could tell which of the groups is correct, if any.

\begin{figure}[htb]
	\includegraphics[width=0.95\columnwidth]{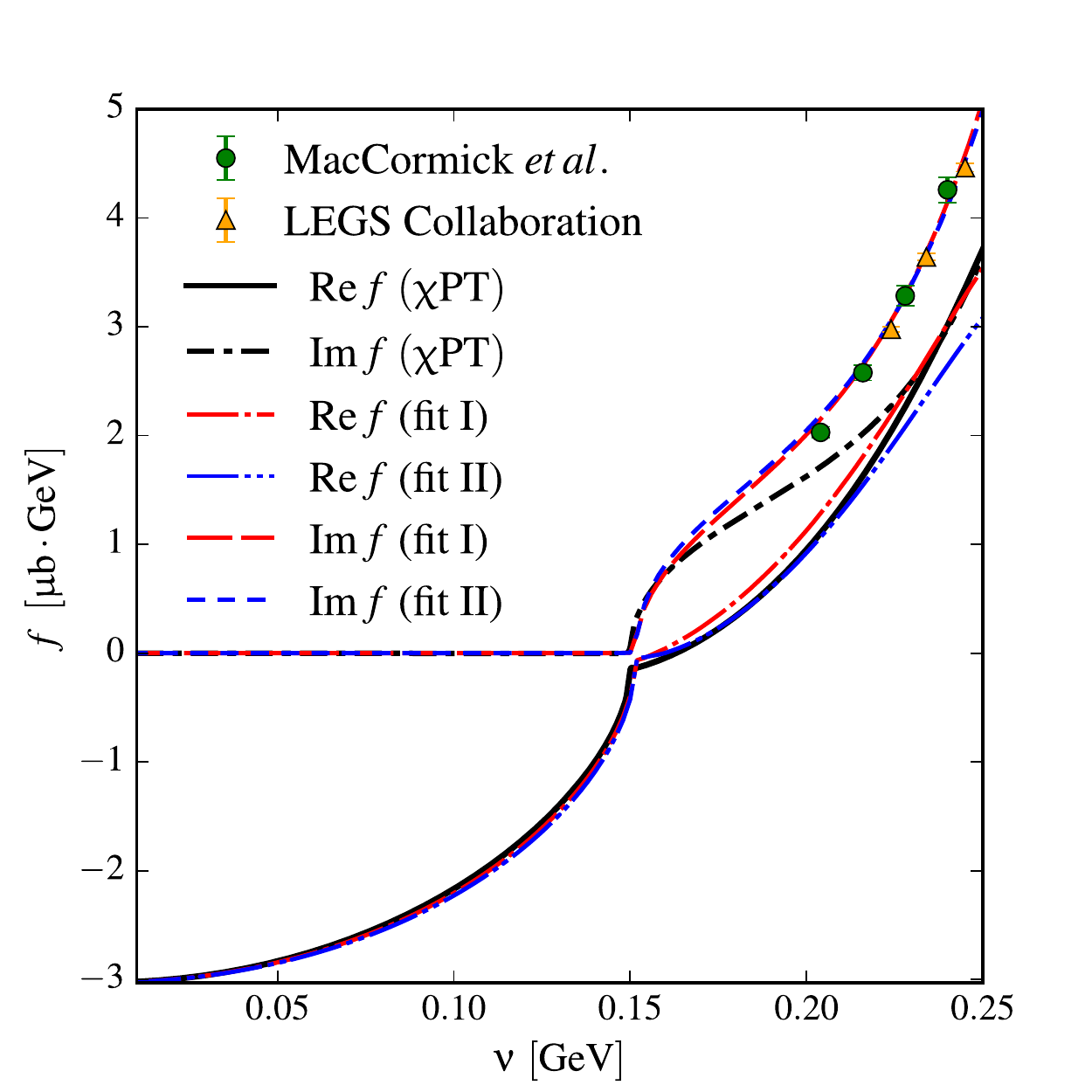}
	\caption{(Color online) Our evaluations of $f(\nu)$ compared to the $\chi$PT calculation
of Ref.~\cite{Lensky:2009uv}.}
	\label{fig:ReIm_f}
\end{figure}

\begin{figure}[hbt]
	\includegraphics[width=0.92\columnwidth]{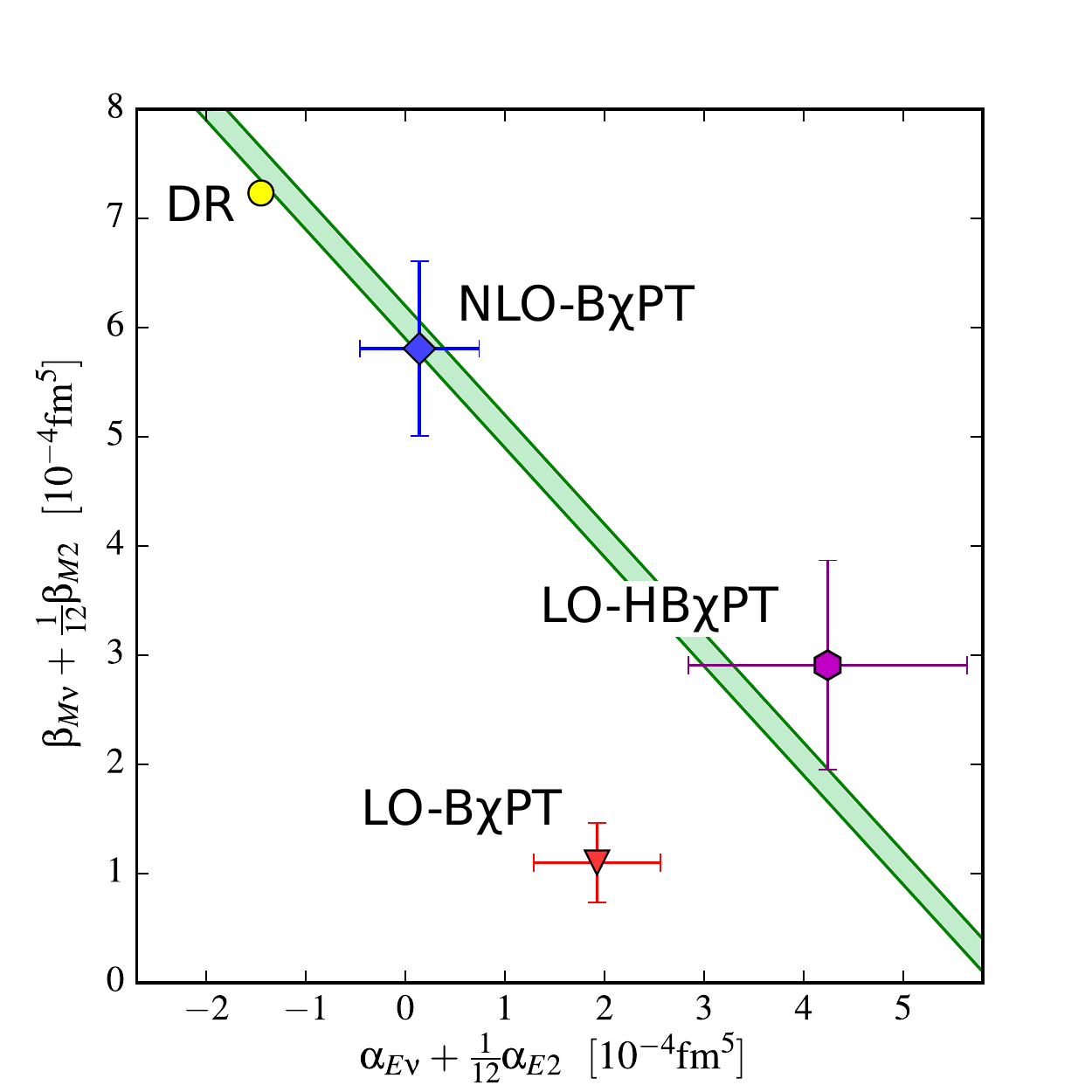}
	\caption{(Color online) The 4\textsuperscript{th}-order sum rule constraint for $\alpha_{E\nu} +\, \frac{1}{12} \alpha_{E2}$ and $\beta_{M\nu} +\, \frac{1}{12} \beta_{M2}$ combinations of polarizabilities, compared to results from dispersion relation approaches (DR) \cite{Babusci:1998ww,Drechsel:2002ar}, baryon chiral perturbation theory (B$\chi$PT) \cite{Lensky:2015}, and heavy baryon chiral perturbation theory (HB$\chi$PT) \cite{Holstein:1999uu}. The errors of the $\chi$PT derive from our  crude estimate of the next-order corrections.}
	\label{fig:4thSR}
\end{figure}

Figure~\ref{fig:ReIm_f} shows both the real and imaginary
part of $f$ at lower energies, where it can be compared
with a calculation done within chiral perturbation theory ($\chi$PT) \cite{Lensky:2009uv}. A rather nice agreement between theory and empirical evaluations is observed for energies up to
about the pion-production threshold.

For very low energy this comparison can
be made more quantitative by looking at the polarizabilities. While for the Baldin
sum rule the situation was extensively discussed in the literature (cf.~\cite{Griesshammer:2012we} for a recent review), the 
4\textsuperscript{th}-order sum rule was not studied at all. 
It can,
however, be very useful in unraveling the higher-order polarizabilities,
as illustrated by \Figref{4thSR}. This is the plot of a combination of proton magnetic
polarizabilities versus electric, where the various theory predictions are compared with our 4\textsuperscript{th}-order sum rule evaluation. The band representing the sum rule covers the 
interval between the two values given in 
Table~\ref{sumruletest} (rows `fit I' and `fit II'). The sum rule clearly provides a
model-independent constraint on these polarizabilities and a rather stringent test for the theoretical approaches.

% \begin{figure}
%   \includegraphics[width=\columnwidth]{246_hw.pdf}
% 	\caption{
% 		(Color online)
% 		Values of Baldin (2nd order), 4\textsuperscript{th} and 6th order sum rules versus integral cut-off energy. \textit{Solid} lines represent values obtained with fit I, while \textit{dashed} lines represent values obtained with fit II. Asymptotic values of integrals are displayed with triangles on the right.}
% 	\label{fig:B46}
% \end{figure}

\section{Conclusion}
\label{sec:summary}
The fundamental relation between the photon absorption and scattering, encompassed in the Kramers-Kronig type of relations, allows us
to evaluate the forward Compton scattering off protons 
using the empirical knowledge of the total photoabsorption cross sections.
The present database of the unpolarized photoabsorption cross section is not entirely
consistent and so as to reflect that we obtain two distinct fits to it.
The two fits yield slightly different results for the spin-independent
amplitude $f(\nu)$, and hence for its low-energy expansion characterized
by the scalar  polarizabilities of the proton. Our two results for the
sum of dipole polarizabilities (or, Baldin sum rule) correspond nicely
with the  results of previous evaluations, which too can be separated into
two groups: 
the {\it old} \cite{Damashek,Schroder}, with the value slightly above 14 (in units of $10^{-4}$ fm\textsuperscript{3}), and 
the {\it new}~\cite{Babusci,Olmos}, with the value slightly below 14.
The 1996 DAPHNE@MAMI experiment \cite{MacCormick}, superseding
the 1972 experiment of 
Armstrong {\it et al.}~\cite{Armstrong}, is clearly responsible
for this difference. Neglecting the older data in favor of the newer ones, yields the lower value of the Baldin sum rule, and vice versa. 
While one can take a preference in one of the two fits and corresponding
results, we prefer to think of their difference
as a systematic uncertainty in the present 
evaluation of the polarizabilities  and of the forward spin-independent amplitude of the proton. 

As far as polarizabilities are concerned,
only the Baldin sum rule is appreciably affected by the inconsistency
in the photoabsorption database. Nevertheless, the two results
(fit I and II in Table~\ref{sumruletest}) are not in conflict with
each other,
given the overlapping error bars. It is customary to take a statistical average in such cases.
Taking a weighted average\footnote{For the weighted average, $\bar x\pm \bar \si$, over 
a set $\{ x_i \pm \si_i \}$, we use \cite[p.~120]{Parratt:1961}: 
$$\bar x = \frac{\sum_i x_i/\si_i^2}{\sum_j 1/\si_j^2},\; 
\bar \si = \Big(\frac{\sum_i (x_i-\bar x)^2/\si_i^2}{\sum_j 1/\si_j^2}\Big)^{1/2}.$$ } over our two values for the Baldin sum rule  we obtain:
$\al_{E1}^{(\mathrm{p})}+\beta_{M1}^{(\mathrm{p})}=(14.0\pm 0.2)\times 10^{-4}\,\mathrm{fm}^3.$
The error bar here does directly not include
the aforementioned systematic uncertainty of the cross section database.
However, since the two results are fairly well surmised by the weighted
average, the latter should be less prone to 
the systematic uncertainty of the database.

We have presented a first study of the sum rule
involving the quadrupole polarizabilities, \Eqref{4thSR},
here referred to as 'the 4\textsuperscript{th}-order sum rule'. 
Our weighted average value for this sum rule, in the proton case, 
is $6.04(4) \times 10^{-4}\,\mathrm{fm}^5$. It agrees
very nicely with the state-of-the-art calculations 
of these polarizabilities based
on fixed-$t$ dispersion relations and chiral perturbation theory,
see \Figref{4thSR}. We note that, while the calculations
demonstrate significant differences in the values of individual
higher-order polarizabilities, these differences apparently 
cancel out from the
forward combination of these polarizabilities which enters the sum rule.

In the subsequent paper we will discuss the evaluation  of the forward {\it spin-dependent} amplitude
$g(\nu)$ and  related sum rules for the forward spin polarizabilities
of the proton. The knowledge of the two amplitudes will allow us
to reconstruct the observables for the proton Compton scattering
at zero angle.

\begin{acknowledgments}
We thank J\"urgen Ahrens for kindly supplying us with a database of total photoabsorption cross sections.
This work was supported by the Deutsche Forschungsgemeinschaft (DFG) through the Collaborative Research Center SFB 1044 [The Low-Energy Frontier of the Standard Model], and the Graduate School DFG/GRK 1581
[Symmetry Breaking in Fundamental Interactions].
\end{acknowledgments}

\onecolumngrid
\appendix

%\section{Compton scattering in scalar QED to one loop}
\section{Sum rules for elastic contribution in scalar QED}
\seclab{scalarQEDExample}

\twocolumngrid
Consider the elastic forward scattering of a photon with momentum
$q$ from a  charged spinless particle with four-momentum $p$ and mass $M$. In the forward direction ($t=0$) this process is completely described by a single amplitude $f(\nu)$. The tree-level QED calculation
(\Figref{TreeLevelCS}) yields immediately $f^{(1)}(\nu) = -\al/M$, where  
we have chosen the normalization of this amplitude to coincide 
with the analogous amplitude for the spin-1/2 case [see \Eqref{definition}]; the  
superscript indicates the order of $\al$.

\begin{figure}[h]
  \centering
    \includegraphics[width=0.28\columnwidth]{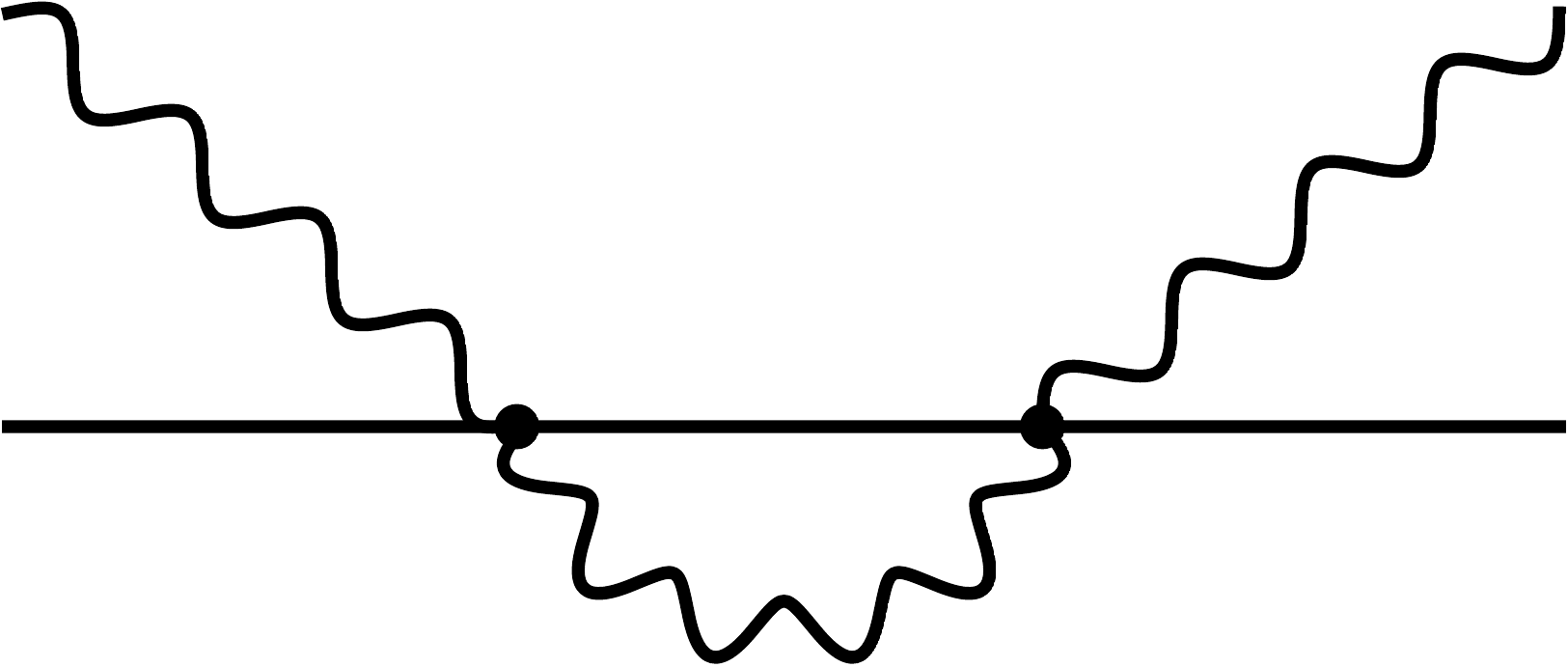} \hskip0.2\columnwidth 
    \includegraphics[width=0.28\columnwidth]{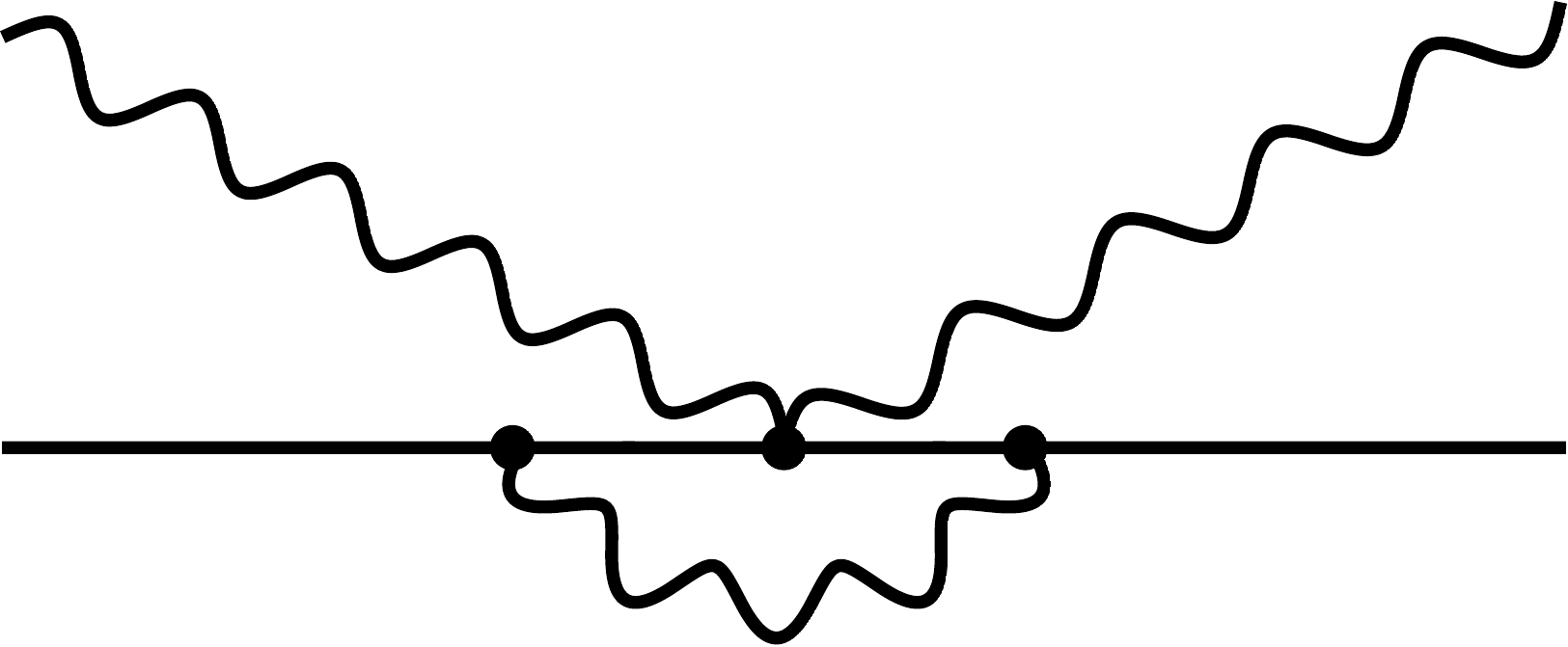}
    \includegraphics[width=0.28\columnwidth]{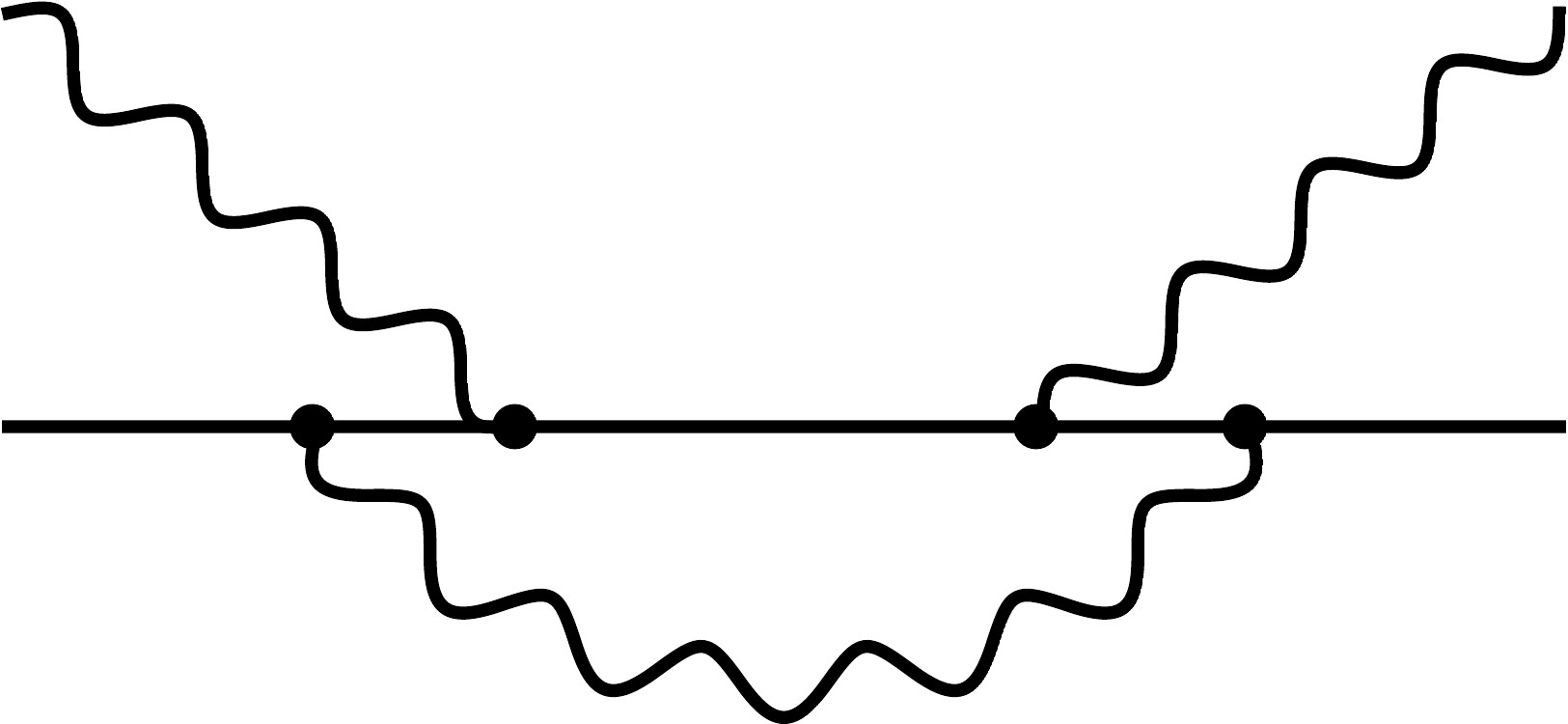}\\[1mm]
    \includegraphics[width=0.28\columnwidth]{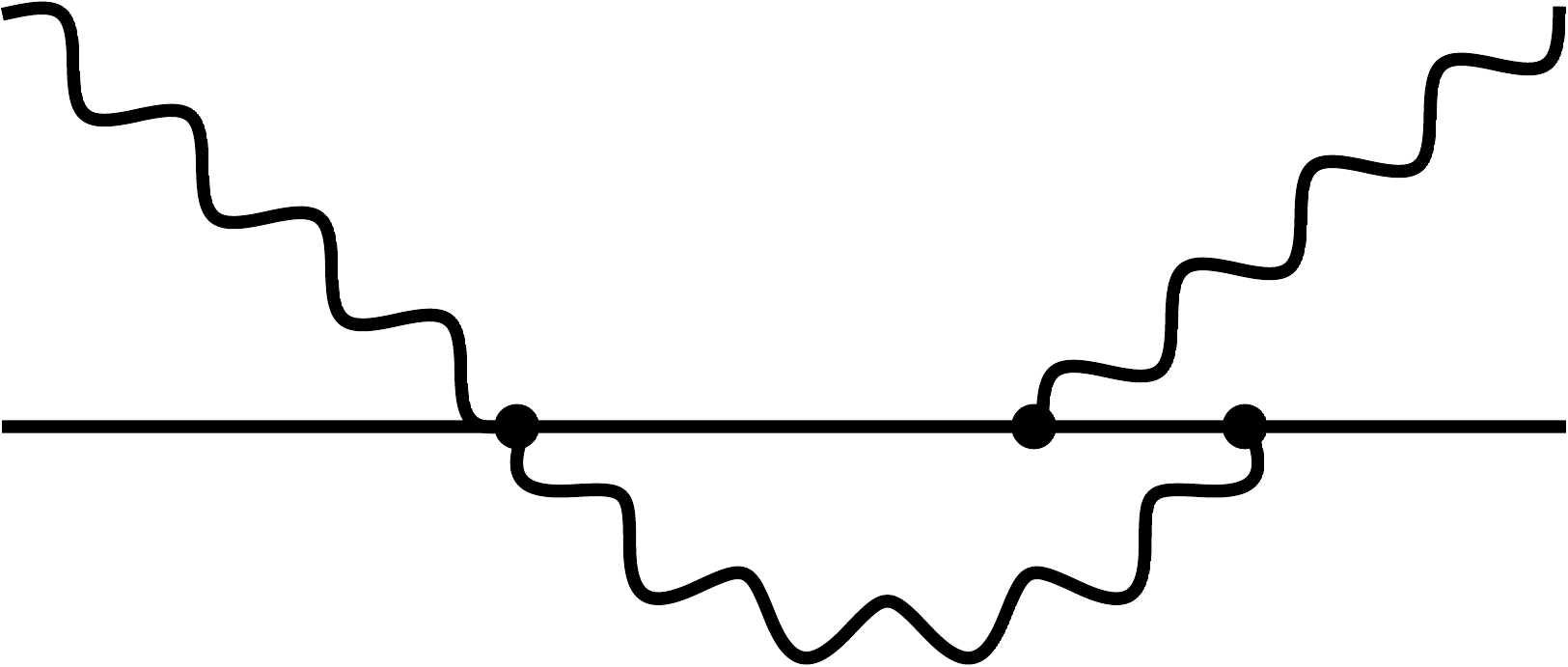}
    \hskip0.2\columnwidth
    \includegraphics[width=0.28\columnwidth]{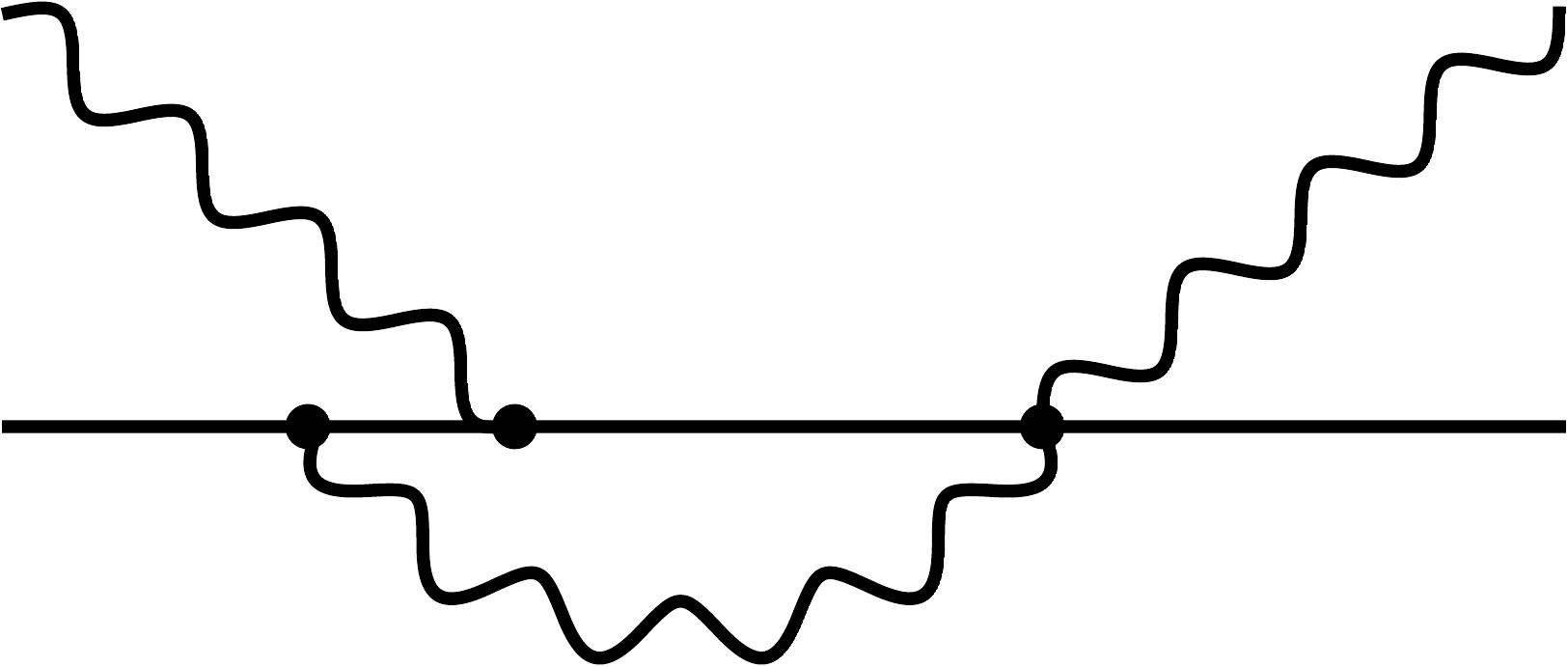}\\[1mm]
    
	\caption{One-loop graphs contributing to the forward CS. Diagrams obtained from these by crossing of the photon lines are included too.}  
	\label{IMdiagrams}
\end{figure}

Next we consider the one-loop corrections. Figure \ref{IMdiagrams} shows the one-particle-irreducible (1PI) diagrams appearing in  scalar QED. The corresponding one-particle-reducible (1PR) diagrams vanish in forward direction, due to the transversality of the photon polarization vector $\eps$ with respect to any of the four momenta, i.e.: $q\cdot\epsilon=0=p\cdot\epsilon$.

Renormalization of these 
diagrams amounts to subtracting their contribution at $\nu=0$. 
We thus find the following expression for the
renormalized amplitude at order $O(\al^2)$:
\begin{figure}[h]
  \centering
  \includegraphics[width=0.26\columnwidth]{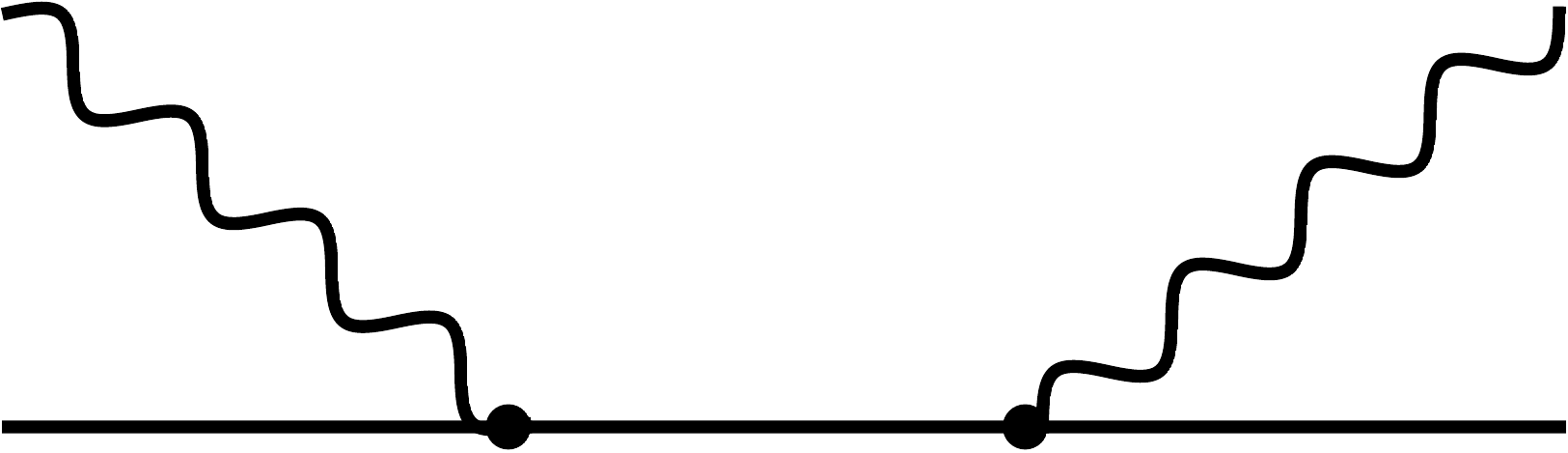} \hskip0.1\columnwidth 
    \includegraphics[width=0.26\columnwidth]{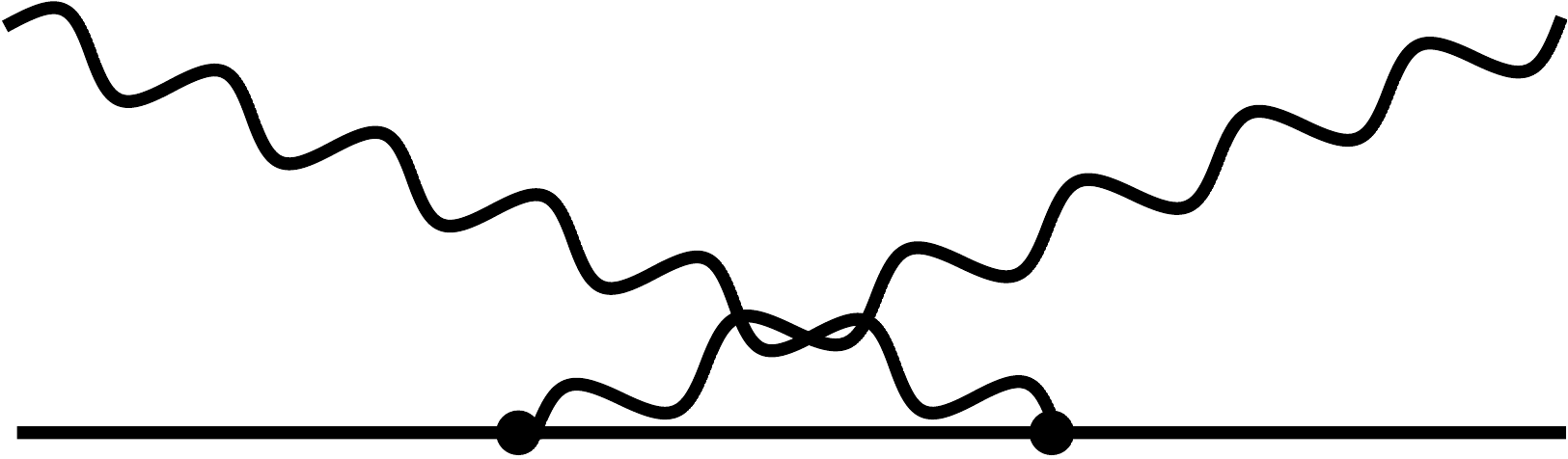} \hskip0.1\columnwidth 
    \includegraphics[width=0.26\columnwidth]{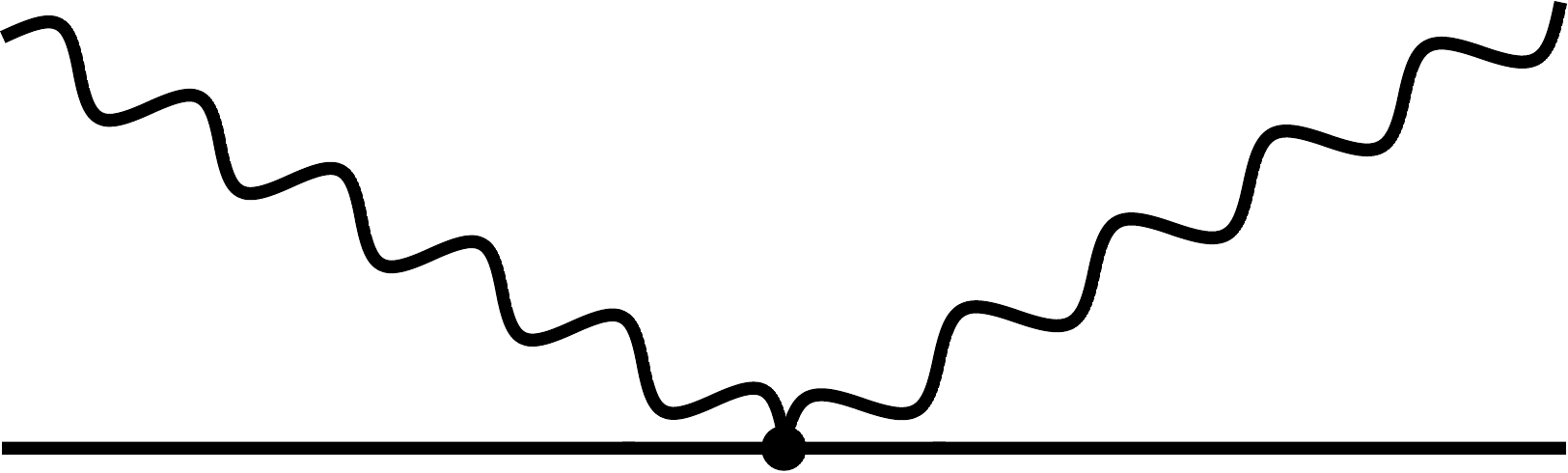}
  	\caption{Tree-level CS diagrams.}  
	\figlab{TreeLevelCS}
\end{figure} 

\begin{widetext}
\bea
\label{eq:scalarscatteringamp}
f^{(2)}(\nu)
&=& \frac{\alpha^2 }{2\pi M} \Bigg\{ \frac{\pi^2 M(M -\nu)+12 \nu ^2}{6 \nu ^2}+\frac{8 \nu ^2 }{\left(M^2-4 \nu ^2\right)}\ln \frac{2 \nu }{M}+\frac{M(M+\nu)}{\nu^2}\ln \frac{2 \nu }{M}\ln \left(1+\frac{2 \nu }{M}\right)\nn\\
&+&\frac{M}{\nu ^2} \left[ (M+\nu ) \,\mathrm{Li}_2\Big(-\frac{2 \nu }{M}\Big)- (M-\nu ) \,\mathrm{Li}_2\Big(1-\frac{2 \nu }{M}\Big)\right] \Bigg\}+\frac{i}{4\pi}\nu\sigma^{(2)}(\nu),
\eea
where Li$_2(x)$ is the dilogarithm, and $\sigma^{(2)}(\nu)$ is the total CS cross section arising at the tree level (cf.~\Figref{TreeLevelCS}):
\begin{align}
\label{eq:QED0xsn}
\si^{(2)}(\nu) =&\frac{2\pi \al^2}{\nu^2} \left\{\frac{2(M+\nu)^2}{M^2+2M\nu}
  -\Big(1+\frac{M}{\nu} \Big) \ln\Big(1+\frac{2\nu}{M}\Big) \right\}.
\end{align}
%\end{widetext}
We note that in the low-energy limit it reproduces the Thomson cross section:
 $ \si^{(2)}(0)  = 8\pi \al^2/3M^2$,
a result that is unaltered by loop corrections, i.e.\ $\si(0) = \si^{(2)}(0)$.

As the total photoabsorption cross section, to this order in $\al$, is given entirely
by the tree-level CS cross section, the fact that $\im f^{(2)} (\nu)=\nu\,\sigma^{(2)}(\nu)/4\pi $ coincides in this case with the statement of the optical theorem. 
We have also checked that the one-loop amplitudes
satisfies the once-subtracted dispersion relation:
\begin{equation}
\eqlab{scalarSR}
f^{(2)}(\nu)=\frac{\nu^2}{2\pi^2}\int_0^\infty \! \dd\nu' \,\frac{\sigma^{(2)}(\nu')}{\nu^{\prime \, 2}-\nu^2 - i0^+},
\end{equation}
and hence the full amplitude, $f^{(1)}+f^{(2)}$, indeed enjoys   the Kramers-Kronig relation given in
\Eqref{KK}.

Now, the whole point of this exercise is to understand the low-energy expansion, and
thus the polarizability sum rules, in the case when the photoabsorption cross section is not vanishing at $\nu=0$.
Expanding the real part of \Eqref{scalarSR} around $\nu=0$,  we find:
%\begin{widetext}
\bea 
\eqlab{scalarLEX}
 \frac{\alpha^2}{\pi M} \bigg(\frac{1 +24 \ln \frac{2\nu}{M}}{9 M^2}\,\nu^2
+\frac{8 (14 +330 \ln \frac{2\nu}{M})}{225 M^4}\,\nu^4
+\frac{4 (17 +616 \ln \frac{2\nu}{M})}{49 M^6}\,\nu^6   + \ldots \bigg)
&=& \frac{1}{2\pi^2} \sum_{n=1}^\infty \nu^{2n} 
\int_0^\infty \!\dd\nu' \,\frac{\sigma^{(2)}(\nu')}{\nu^{\prime\, 2n}}.
\eea 
%\end{widetext}
Hence, the coefficients diverge in the infrared. However, there is an apparent mismatch:  
they are logarithmically divergent on one side
 and power-divergent on the other. To match the sides exactly at
 each order of $\nu$, thus defining the sum rules for ``quasi-static'' 
 polarizabilities, we subtract all the power divergences on the
 right-hand side (rhs) and 
regularize both sides with the same infrared cutoff (equal to $\nu$):
%\begin{widetext}
\bea 
\eqlab{goodscalarLEX}
 \frac{\alpha^2}{\pi M} \bigg(\frac{1 +24 \ln \frac{2\nu}{M}}{9 M^2}\,\nu^2
+\frac{8 (14 +330 \ln \frac{2\nu}{M})}{225 M^4}\,\nu^4  + \ldots \bigg)
&=& \frac{1}{2\pi^2} \sum_{n=1}^\infty \nu^{2n} 
\int_\nu^\infty \!\dd\nu' \,\frac{\sigma^{(2)}(\nu')-\sum\limits_{k=0}^{2(n-1)}\frac{1}{k!}\frac{\dd^k\sigma^{(2)}(\nu)}{\dd\nu^k}\Big|_{\nu=0}\,\nu^{\prime\, k}}{\nu^{\prime\, 2n}}.\nn\\
\eea  
\end{widetext}
Both sides are  now identical at each order of $\nu$. This is nontrivial, at least for the analytic terms; the logs are fairly easily obtained from the non-regularized right-hand side (rhs) in \Eqref{scalarLEX}, cf.~\cite{Holstein:2005db}.

Extending these arguments to all orders in $\al$, we find that the proper
low-energy expansion for the `elastic' part of the amplitude [see \Eqref{fel}]
reads as:
\beq
\eqlab{elLEX}
 f_{\mathrm{el}}(\nu)= -\,\frac{\al}{M} + \frac{1}{2\pi^2}\sum_{n=1}^\infty \nu^{2n} 
\!\int_\nu^\infty\! \dd\nu' \,\frac{\sigma(\nu')-\bar\sigma_n(\nu')}{\nu^{\prime\, 2n}}\,,
\eeq
where $\si $ is the total cross-section of Compton scattering and
$\bar\sigma_n$ are the infrared subtractions:
\beq 
\bar\sigma_n(\nu') \equiv \sum\limits_{k=0}^{2(n-1)}\frac{1}{k!}\frac{\dd^k\sigma(\nu)}{\dd\nu^k}\Big|_{\nu=0}\,\nu^{\prime\, k}.
\eeq 

Now we can, for instance, formulate the Baldin sum rule for the
elastic contribution to the dipole polarizabilities. By definition
\beq
f_{\mathrm{el}}(\nu) = -\al/M + (\alpha_{E1}+\beta_{M1})_{\mathrm{el}}\;\nu^2 + O(\nu^4),
\eeq
and hence, matching it with the rhs of \Eqref{elLEX}, we obtain:
\begin{equation}
\label{scalarBSR}
(\alpha_{E1}+\beta_{M1})_{\mathrm{el}} =\frac{1}{2\pi^2}\int_\nu^\infty \dd\nu' \,\frac{\sigma(\nu')-\sigma(0)}{\nu^{\prime\,2}}.
\end{equation}
In our scalar QED example, where $\si$ is the tree-level cross section $\si^{(2)}$, we obtain 
\begin{equation}
\big(\alpha_{E1}^{(2)}+\beta_{M1}^{(2)}\big)_{\mathrm{el}}=\frac{\alpha^2}{9\pi M^3 }\left( 1+24 \ln \frac{2\nu}{M}\right),
\end{equation}
which of course reproduces the
one-loop result [cf.~the first term in the expansion 
of $f^{(2)}$ in \Eqref{goodscalarLEX}].

\end{document}